\documentclass[pdflatex,prl,twocolumn,showpacs,superscriptaddress,preprintnumbers]{revtex4-2}

\usepackage{bm}
\usepackage{color}
\usepackage{xcolor}
\usepackage{soul}
\usepackage{physics}
\usepackage{placeins}
\usepackage{amsmath,amssymb}
\usepackage{graphicx,multirow,tabularx}
\usepackage[ruled,vlined]{algorithm2e}
\usepackage{lineno}
\usepackage[colorlinks=true,pdfstartview=FitV,linkcolor=blue,citecolor=magenta,urlcolor=blue,bookmarks=true,bookmarksnumbered=true]{hyperref}

\newcommand{\Sec}[1]{\textit{#1.}---}

\begin{document}
%\linenumbers
\title{Locating Critical Points Using Ratios of Lee-Yang Zeros}

\author{Tatsuya Wada}
\email{tatsuya.wada@yukawa.kyoto-u.ac.jp}
\affiliation{
  Yukawa Institute for Theoretical Physics, Kyoto University, Kyoto, 606-8502, Japan}
\author{Masakiyo Kitazawa}
\affiliation{
  Yukawa Institute for Theoretical Physics, Kyoto University, Kyoto, 606-8502, Japan}
\affiliation{
  J-PARC Branch, KEK Theory Center, 
  Institute of Particle and Nuclear Studies, KEK, Tokai, Ibaraki 319-1106, Japan}
\author{Kazuyuki Kanaya}
\affiliation{
  Tomonaga Center for the History of the Universe, University of Tsukuba, Tsukuba, Ibaraki 305-8571, Japan}
\date{\today}
\preprint{YITP-24-138, J-PARC-TH-0309, UTHEP-792}
\begin{abstract}
We propose a method to numerically determine the location of a critical point in general systems using the finite-size scaling of Lee-Yang zeros. This method makes use of the fact that the ratios of Lee-Yang zeros on various spatial volumes intersect at the critical point. While the method is similar to the Binder-cumulant analysis, it is advantageous in suppressing the finite-volume effects arising from the mixing of variables in general systems. We show that the method works successfully for numerically locating the CP in the three-dimensional three-state Potts model with a nonzero external field.
\end{abstract}

\maketitle

\Sec{Introduction}
Critical points (CPs) are interesting research objects in physics, which appear in various systems from water~\cite{Water} to nuclear matter~\cite{Pochodzalla:1995xy} that are separated about ten orders of magnitude in temperature. Although the existence and location of a CP are not constrained by symmetries of the system, once a CP manifests itself in a system the thermodynamic properties around it are tightly restricted by the scaling law and universality class~\cite{Goldenfeld,Nishimori}. These properties are not only intriguing research subjects in statistical mechanics but also useful tools for revealing phase transitions in nontrivial systems. For example, the scaling properties have been actively utilized in the numerical and experimental analyses of the chiral phase transition and conjectured CPs in quantum chromodynamics (QCD)~\cite{Parotto:2018pwx,Philipsen:2021qji,HotQCD:2019xnw,Ratti:2020oro,Ding:2024sux,Asakawa:2015ybt,Bzdak:2019pkr}. 

In an investigation of a CP in numerical simulations, it is crucial to properly deal with the finite-volume effects since the simulations are always performed on finite volumes. It is known that thermodynamics in the vicinity of a CP on finite but sufficiently large volume obey the finite-size scaling (FSS)~\cite{Pelissetto:2000ek,Binder:2001ha}. The scaling properties of various susceptibilities, i.e. quantities given by derivatives of the free energy, obtained from the FSS have been used in the simulations for determining properties of the CP, such as its location and critical exponents~\cite{Binder:1981sa,Pelissetto:2000ek,Binder:2001ha,Ferrenberg:2018zst,Kaupuzs:2022wtd}. 

In the present Letter, we focus on the use of the FSS of Lee-Yang zeros (LYZ)~\cite{Yang:1952be,Lee:1952ig}, i.e. zeros of the partition function on the complex-variable space, for numerical investigations of a CP in general systems. Whereas the FSS of LYZ has been investigated in the literature~\cite{Kenna:1993fp,PhysRevE.56.2418,BENA_2005,Deger_2020,DEmidio:2023tsn}, to the best of our knowledge its systematic utilization for this purpose has not been discussed so far. The LYZ have been investigated in lattice-QCD numerical simulations~\cite{Fodor:2004nz,Ejiri:2005ts,Nagata:2014fra,Giordano:2019slo,Giordano:2019gev}. In particular, its application for locating the CP in QCD at nonzero chemical potential has been discussed recently~\cite{Dimopoulos:2021vrk,Basar:2023nkp,Zambello:2023ptp,Clarke:2024ugt} in connection to the Lee-Yang edge singularity (LYES)~\cite{Stephanov:2006dn,Ejiri:2014oka,An:2017brc,Basar:2021gyi,Rennecke:2022ohx,Johnson:2022cqv,Singh:2023bog,Karsch:2023rfb}. However, the finite-volume effects have not been investigated in detail in these studies. As we discuss below, systematic utilization of the FSS of LYZ provides us with a general procedure applicable to a wide variety of numerical simulations including lattice QCD and those in statistical physics.

%As a systematic procedure to study a CP in numerical simulations, based on the FSS 
%w
We show that the ratios of the imaginary parts of LYZ obtained on different volumes intersect at the CP in the large-volume limit. We propose the use of this property, which is similar to the Binder cumulants~\cite{Binder:1981sa}, for locating a CP in numerical simulations. The LYZ carry information of the system that is not encoded in finite derivatives of the free energy. Our method thus serves as a %novel 
procedure independent of the ordinary methods that rely on susceptibilities. We test the method for analyzing the CP in the three-dimensional three-state Potts model and show that the method can determine its location successfully with almost the same statistical error compared to the Binder-cumulant method. As byproducts, %some useful properties of LYZ, especially their relation to the LYES, will also be clarified. 
we also derive some useful properties of LYZ, especially their relation to the LYES, which play essential roles in controlling finite-size effects in numerical results.

\Sec{Ising model}
To illustrate the method, we start from a simple case of the conventional three-dimensional Ising (3D-Ising) model~\cite{Goldenfeld} described by the reduced temperature $t$ and external magnetic field $h$ on the cubic lattice of size $L^3$, having a CP at $(t,h)=(0,0)$ and a first-order phase transition at $h=0$ for $t<0$. % for $L\to\infty$. 
We denote the partition function of this model as $Z(t,h,L^{-1})$, which satisfies $Z(t,h,L^{-1})=Z(t,-h,L^{-1})$.
The LYZ of this model are the values of $h\in\mathbb{C}$ satisfying $Z(t,h,L^{-1})=0$ for a given $t\in\mathbb{R}$~\cite{Yang:1952be,Lee:1952ig}. It is known as the Lee-Yang circle theorem that the LYZ distribute discretely on the pure-imaginary axis for finite $L$~\cite{Lee:1952ig}. In the following, we denote the LYZ with ${\rm Im}\ h>0$ as $h=h_{\rm LY}^{(n)}(t,L)$, where $n=1,2,\cdots$ labels different LYZ so that $0<{\rm Im}\ h_{\rm LY}^{(1)}(t,L)<{\rm Im}\ h_{\rm LY}^{(2)}(t,L)<\cdots$; the system has the other LYZ at $h=-h_{\rm LY}^{(n)}(t,L)$. Since $Z(t,h,L^{-1})$ at finite $L$ is a regular function of $t$ and $h$, $h_{\rm LY}^{(n)}(t,L)$ should also be regular functions of $t$ for finite $L$~\footnote{The regularity of $h_{\rm LY}^{(n)}(t,L)$ is violated when $Z(t,h,L^{-1})=0$ has a multiple solution at some $t$ even for a regular $Z$. We, however, have numerically checked that such cases do not occur in the parameter range explored in this study.}.

According to the FSS, the partition function in the vicinity of the CP for different $L$ is represented by the scaling function $\tilde Z(\tilde t,\tilde h)$ as 
\begin{align}
    Z(t,h,L^{-1}) 
    = \tilde{Z}(L^{y_t}t,L^{y_h}h) ,
    \label{eq:FSS}
\end{align}
for sufficiently large $L$. In the 3D-Ising model, the values of the exponents, $y_t \simeq 1.588$ and $y_h \simeq 2.482$, have been analyzed with high precision~\cite{Kos:2016ysd,Ferrenberg:2018zst,Kaupuzs:2022wtd}.
Since the LYZ are given by zeros of Eq.~\eqref{eq:FSS}, it immediately follows that the LYZ for different $L$ obey~\cite{ITZYKSON1983415}
\begin{align}
    L^{y_h} h_{\rm LY}^{(n)}(t,L) = \tilde{h}_{\rm LY}^{(n)}(L^{y_t}t) ,
    \label{eq:tildeh_LY}
\end{align}
with $\tilde{h}_{\rm LY}^{(n)}(\tilde t)$ satisfying $\tilde Z(\tilde t,\tilde{h}_{\rm LY}^{(n)}(\tilde t))=0$.

For $t<0$ and $L\to\infty$, the LYZ are densely distributed around $h=0$ reflecting the discontinuity of the first-order phase transition~\cite{Lee:1952ig}, which means that $h_{\rm LY}^{(n)}(t,L)\to0$ in this limit for finite $n$. For $t>0$, since $\lim_{L\to\infty}Z(t,h,L^{-1})$ is a regular function at $h=0$, the distribution of LYZ for $L\to\infty$ terminates at nonzero (pure-imaginary) values of $h$ away from $h=0$, which is called the LYES~\cite{Kortman:1971zz}. Denoting the LYES as $h=\pm h_{\rm LYES}(t)$ we obtain
\begin{align}
    h_{\rm LY}^{(n)}(t,L)=L^{-y_h} {\tilde h}_{\rm LY}^{(n)}(L^{y_t}t) \xrightarrow[L\to\infty]{} h_{\rm LYES}(t) 
    \quad
    (t>0),
    \label{eq:LYES}
\end{align} 
for finite $n$. Since the right-hand side of Eq.~\eqref{eq:LYES} does not depend on $L$, only a possible asymptotic behavior of ${\tilde h}_{\rm LY}^{(n)}({\tilde t})$ for $\tilde t\to\infty$ is ${\tilde h}_{\rm LY}^{(n)}({\tilde t})\propto {\tilde t}^{y_h/y_t}$, which yields $h_{\rm LYES}(t)\propto t^{y_h/y_t}$ for $t>0$~\cite{BENA_2005}.

\begin{figure}
    \centering
\includegraphics[width=0.47\linewidth]{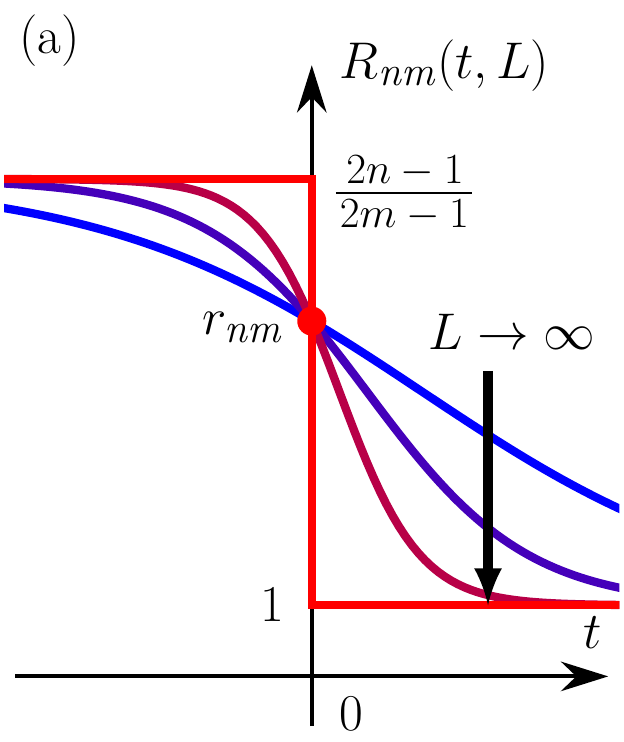}
\hspace{0.01\textwidth}
\includegraphics[width=0.47\linewidth]{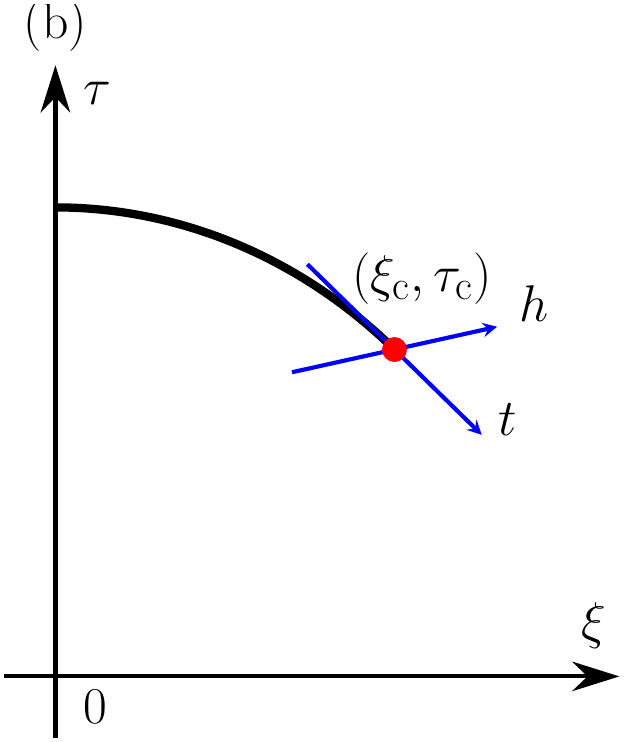}
    \caption{(a) Schematic behavior of $R_{nm}(t,L)$ for different $L$. (b) Phase diagram of the three-dimensional three-state Potts model~\eqref{eq:H_Potts}. The black-solid line shows the first-order phase transition that terminates at a CP denoted by the red circle. The Ising variables $t$ and $h$ are encoded as blue arrows.}
    \label{fig:cartoon}
\end{figure}

Now we focus on the ratio of two LYZ at the same $L$
\begin{align}
    R_{nm}(t,L) = \frac{h_{\rm LY}^{(n)}(t,L)}{h_{\rm LY}^{(m)}(t,L)}
    = \frac{\tilde h_{\rm LY}^{(n)}(L^{y_t}t)}{\tilde h_{\rm LY}^{(m)}(L^{y_t}t)} .
    \label{eq:Rnm}
\end{align}
Near $t=0$, from the regularity $\tilde{h}_{\rm LY}^{(n)}(\tilde t)$ is Taylor expanded as 
\begin{align}
    \tilde{h}_{\rm LY}^{(n)}(\tilde t) = i\big( X_{n} + Y_{n}\tilde t + {\cal O}(\tilde{t}\,^2) \big) ,
    \label{eq:XY} 
\end{align}
with real numbers $X_n$ and $Y_n$.
Substituting Eq.~\eqref{eq:XY} into Eq.~\eqref{eq:Rnm} one obtains 
\begin{align}
    & R_{nm}(t,L) = r_{nm} + c_{nm} L^{y_t} t + {\cal O}(t^2) ,
    \label{eq:Rlinear}
\end{align}
with $r_{nm} = X_n/X_m$ and $c_{nm} = r_{nm} (Y_n/X_n - Y_m/X_m )$. Equation~\eqref{eq:Rlinear} shows that $R_{nm}(0,L)=r_{nm}$ is independent of $L$, while the slope at $t=0$ scales as $L^{y_t}$. In other words, $R_{nm}(t,L)$ for different $L$ intersects at $t=0$ as in Fig.~\ref{fig:cartoon}~(a). This property is similar to the Binder cumulants~\cite{Binder:1981sa} and would provide an alternative method to determine the critical temperature from the intersection point in numerical simulations. 

It is also shown that Eq.~\eqref{eq:Rnm} for $L\to\infty$ behaves as
\begin{align}
    R_{nm}(t) \xrightarrow[L\to\infty]{}
    \begin{cases}
        \frac{2n-1}{2m-1} & (t<0) \\
        1 & (t>0)
    \end{cases}
    \qquad
    (\mbox{finite}~ n,m),
    \label{eq:Rlim}
\end{align}
(see, Fig.~\ref{fig:cartoon}~(a)).
First, Eq.~\eqref{eq:Rlim} for $t<0$ is obtained from the fact that the LYZ for sufficiently large $L$ are aligned with an equal distance as $h_{\rm LY}^{(n)}(t,L)=a(t)(2n-1)/L^3$ with a pure-imaginary function $a(t)$~\cite{Ejiri:2005ts}. Second, Eq.~\eqref{eq:Rlim} for $t>0$ follows from Eq.~\eqref{eq:LYES}.

\Sec{CP in general systems}
Next, we extend the argument to CPs in general systems that belong to the same universality class as the 3D-Ising model. We suppose a system whose partition function ${\cal Z}(\tau,\xi,l^{-1})$ is described by two variables $\tau$ and $\xi$ with $l$ being a dimensionless parameter proportional to the spatial size of the system. We also assume that this system has a first-order phase-transition line on the $\tau$--$\xi$ plane that terminates at a CP at $(\tau,\xi)=(\tau_{\rm c},\xi_{\rm c})$ as depicted in Fig.~\ref{fig:cartoon}~(b). 
%In the following, we consider zero points of the partition function for $\xi\in\mathbb{C}$ and $\tau\in\mathbb{R}$ and refer to them as the LYZ $\xi=\pm\xi_{\rm LY}^{(n)}(\tau,l)$, i.e. ${\cal Z}(\tau,\pm\xi_{\rm LY}^{(n)}(\tau,l),l^{-1})=0$, where the definition of the label $n$ is the same as before.
In the following, we refer to zero points of the partition function for $\xi\in\mathbb{C}$ and $\tau\in\mathbb{R}$ as the LYZ, and denote the LYZ for ${\rm Im}\ \xi>0$ as $\xi=\xi_{\rm LY}^{(n)}(\tau,l)$, i.e. ${\cal Z}(\tau,\xi_{\rm LY}^{(n)}(\tau,l),l^{-1})=0$, where the definition of the label $n$ is the same as before.

Since %the CP belongs to the 3D-Ising universality class and there are only two relevant variables at the CP, 
there are only two relevant variables for the CP in the 3D-Ising universality class, the partition function in the vicinity of the CP for large $l$ is related to that of the Ising model as 
\begin{align}
    {\cal Z}(\tau,\xi,l^{-1}) = Z(\check t(\tau,\xi),\check h(\tau,\xi),l^{-1}) ,
    \label{eq:F=F}
\end{align}
where $\check t(\tau,\xi)$ and $\check h(\tau,\xi)$ obey the linear relation
\begin{align}
    \begin{pmatrix}
        \check t \\ \check h
    \end{pmatrix}
    =
    \begin{pmatrix}
        a_{11} & a_{12} \\
        a_{21} & a_{22} 
    \end{pmatrix}
    \begin{pmatrix}
        \tau-\tau_{\rm c} \\ \xi-\xi_{\rm c}
    \end{pmatrix}
    \equiv
    A
    \begin{pmatrix}
        \delta\tau \\ \delta\xi
    \end{pmatrix} .
    \label{eq:A}
\end{align}
Here, the $t$ axis encoded on the $\tau$--$\xi$ plane should be parallel to the first-order phase-transition line at the CP as in Fig.~\ref{fig:cartoon}~(b).

The fact that the LYZ are zeros of Eq.~\eqref{eq:F=F} yields
\begin{align}
    l^{y_h} \check h\big(\tau,\xi_{\rm LY}^{(n)}(\tau,l)\big) 
    = \tilde{h}_{\rm LY}^{(n)}\big(l^{y_t}\check t[\tau, \xi_{\rm LY}^{(n)}(\tau,l)] \big).
    \label{eq:hxi}
\end{align}
Equation~\eqref{eq:hxi} together with Eqs.~\eqref{eq:A} and~\eqref{eq:XY} leads to
\begin{align}
    &l^{y_h} \big[ a_{21} \delta\tau + a_{22} (\xi_{\rm LY}^{(n)}(\tau,l)-\xi_{\rm c}) \big]
    \notag \\
    &= i\big\{ X_n + Y_n l^{y_t} \big[ a_{11} \delta\tau + a_{12} (\xi_{\rm LY}^{(n)}(\tau,l)-\xi_{\rm c}) \big] \big\}+ {\cal O}(\delta\tau^2),
\end{align}
which gives
\begin{align}
    \xi_{\rm LY}^{(n)}(\tau,l)  
    = \xi_{\rm c} + \frac{ iX_n - ( a_{21} l^{y_h} -iY_n a_{11} l^{y_t} )\delta\tau}{ a_{22} l^{y_h} -iY_n a_{12}l^{y_t} },
    \label{eq:xi_LY}
\end{align}
where the terms of order ${\cal O}(\delta\tau^2)$ are suppressed for simplicity.
Using $0<y_t<y_h$ and expanding Eq.~\eqref{eq:xi_LY} by $l^{-1}$ one obtains 
\begin{align}
    &{\rm Re}\ \xi_{\rm LY}^{(n)}(\tau,l)  
    = \xi_{\rm c} - \frac{a_{21}}{a_{22}}\delta\tau + {\cal O}(l^{2\bar y}),
    \label{eq:Rexi} 
    \\
    &{\rm Im}\ \xi_{\rm LY}^{(n)}(\tau,l)  
    = \frac{X_n}{a_{22}} l^{-y_h} + \frac{Y_n\det A}{a_{22}^2} l^{\bar y} \delta\tau + {\cal O}(l^{2\bar y}),
    \label{eq:Imxi}
\end{align}
with $\bar y=y_t-y_h<0$. 
%= (1-\beta\delta)/\nu<0$.

Equation~\eqref{eq:Rexi} shows that $(\tau,{\rm Re}\ \xi_{\rm LY}^{(n)}(\tau,l))$ moves along the $t$-axis with $h=0$ in terms of the Ising variables for $l\to\infty$. Equations~\eqref{eq:Imxi} and~\eqref{eq:LYES} also lead to ${\rm Im}\xi_{\rm LY}^{(n)}(\tau,l)\propto \delta\tau^{y_h/y_t}$ for $l\to\infty$ and $\delta\tau\to0$~\cite{Stephanov:2006dn}. The finite-size corrections to these results are obtained by explicitly calculating the higher-order terms omitted in Eqs.~\eqref{eq:Rexi} and~\eqref{eq:Imxi}. 

To adapt Eq.~\eqref{eq:Rnm} to the present case, we consider the ratios between the {\it imaginary parts} of $\xi_{\rm LY}^{(n)}(\tau,l)$. By expanding them by $\delta\tau$ and $l^{-1}$ one obtains
\begin{align}
    {\cal R}_{nm}(\tau,l) 
    =& \frac{{\rm Im}\ \xi_{\rm LY}^{(n)}(\tau,l)}{{\rm Im}\ \xi_{\rm LY}^{(m)}(\tau,l)}
    \notag \\
    =& \big( r_{nm} + C_{nm} l^{y_t} \delta\tau + {\cal O}(\delta\tau^2) \big) 
    \notag \\
    & \times \big( 1 + D_{nm} l^{2\bar y} + {\cal O}(L^{4\bar y})\big),
    \label{eq:Rlinear2}
\end{align}
where $C_{nm}=c_{nm}\det A /a_{22}$ and $D_{nm}= -(Y_n^2-Y_m^2)a_{12}^2/a_{22}^2$. For $l\to\infty$, Eq.~\eqref{eq:Rlinear2} is dominated by the first bracket on the far-right-hand side. This means that the intersection point of Eq.~\eqref{eq:Rlinear2} converges to the CP as in the Ising model for $l\to\infty$. Notice that $r_{nm}=\lim_{l\to\infty}{\cal R}_{nm}(\tau_{\rm c},l)$ is the same as Eq.~\eqref{eq:Rlinear}, i.e. the value of ${\cal R}_{nm}(\tau,l)$ at the intersection point is unique in individual universality class. For finite $l$, however, the second bracket in Eq.~\eqref{eq:Rlinear2} gives rise to a deviation unless $a_{12}=0$. It is also shown easily that the $l\to\infty$ limit of Eq.~\eqref{eq:Rlinear2} away from $\delta\tau=0$ obeys Eq.~\eqref{eq:Rlim}. 

Now, let us compare Eq.~\eqref{eq:Rlinear2} with the Binder-cumulant method~\cite{Binder:1981sa}. For locating a CP on the $\tau$--$\xi$ plane, one may define the fourth-order Binder cumulant as ${\cal B}_4(\tau,l)={\rm min}_\xi [(\partial^4 {\cal F}(\tau,\xi,l^{-1})/\partial \xi^4)/(\partial^2 {\cal F}(\tau,\xi,l^{-1})/\partial \xi^2)^2]+3$~\cite{Karsch:2000xv,Jin:2017jjp,Kiyohara:2021smr} with the free energy ${\cal F}(\tau,\xi,l^{-1})=-T \ln {\cal Z}(\tau,\xi,l^{-1})$ with temperature $T$. One then obtains~\cite{Jin:2017jjp,Cuteri:2020yke}
\begin{align}
    {\cal B}_4(\tau,l) 
    = \big( b_4 + c_4 l^{y_t} \delta\tau + {\cal O}(\delta\tau^2) \big) 
    %\notag \\ &\times
    \big( 1 + d_4 l^{\bar y} + {\cal O}(l^{2\bar y})\big),
    \label{eq:Rlinear3}
\end{align}
where $d_4$ is proportional to $a_{12}$. Comparing this result with Eq.~\eqref{eq:Rlinear2}, one finds that the second bracket in Eq.~\eqref{eq:Rlinear3} converges slower than that in Eq.~\eqref{eq:Rlinear2} for $l\to\infty$. This implies that the finite-volume effect from $a_{12}\ne0$ is suppressed more quickly for $l\to\infty$ in ${\cal R}_{nm}(\tau,l)$ than ${\cal B}_4(\tau,l)$, which would be an advantage of the former.

\Sec{Numerical analysis in Potts model}
To verify the validity of Eq.~\eqref{eq:Rlinear2} in practical numerical analyses, we perform the Monte-Carlo simulation of the three-dimensional three-state Potts model with Hamiltonian
\begin{align}
    \frac{{\cal H}(\tau,\xi)}{T} = -\tau\sum_{\langle i,j\rangle} \delta_{\sigma_i\sigma_j} - \xi \sum_i \delta_{\sigma_i,1},
    \label{eq:H_Potts}
\end{align}
on the simple cubic lattice of size $L^3$ with periodic boundary conditions, where $\sigma_i$ takes three states $\sigma_i=1,2,3$ with the subscript denoting the lattice site and $\sum_{\langle i,j\rangle}$ represents the sum over all pairs of adjacent sites. 
As schematically shown in Fig.~\ref{fig:cartoon}~(b), this model is $Z(3)$ symmetric and has a first-order phase transition at vanishing external field $\xi=0$,
which is eventually terminated at a CP for $\xi>0$ that belongs to the 3D-Ising universality class~\cite{Karsch:2000xv}. In Ref.~\cite{Karsch:2000xv}, this CP has been investigated by the Binder-cumulant method in relation to the CP in QCD with heavy-mass quarks~\cite{Cuteri:2020yke,Kiyohara:2021smr,Ashikawa:2024njc}.

We generate configurations of Eq.~\eqref{eq:H_Potts} by the heat-bath algorithm for $L=24,30,40,50,60,70$ at three simulation parameters $(\tau_{\rm sim},\xi_{\rm sim})$ near the CP with $\xi_{\rm sim}=0.0007,0.00075,0.0008$ and the corresponding value of $\tau_{\rm sim}$ chosen from Table~I in Ref.~\cite{Karsch:2000xv}. For each parameter, we perform the measurements on $10^6$ configurations separated by ten even-odd heat-bath updates after thermalization.

To numerically search for the LYZ, we use the reweighting method~\cite{PhysRevLett.61.2635,Ejiri:2005ts}, i.e. we search for the zeros of 
\begin{align}
    \frac{{\cal Z}(\tau,\xi,L^{-1})}{{\cal Z}(\tau,\Re\ \xi,L^{-1})}
    = \frac{
    \big\langle e^{-{\cal H}(\tau,\xi)+{\cal H}(\tau_{\rm sim},\xi_{\rm sim})} \big\rangle_L
    }{
    \big\langle e^{-{\cal H}(\tau,\Re\ \xi)+{\cal H}(\tau_{\rm sim},\xi_{\rm sim})} \big\rangle_L
    } ,
    \label{eq:Zreweight}
\end{align}
with $\tau\in\mathbb{R}$ and $\xi\in\mathbb{C}$ for each simulation parameter, where $\langle\cdot\rangle_L$ denotes the average over the configurations at $(\tau_{\rm sim},\xi_{\rm sim})$ of size $L^3$. The numerical cost to calculate Eq.~\eqref{eq:Zreweight} does not depend on $L$ and is negligibly small compared to that for updates of configurations. Whereas the analysis of Eq.~\eqref{eq:Zreweight} suffers from the overlapping problem when $(\tau,\xi)$ is largely deviated from $(\tau_{\rm sim},\xi_{\rm sim})$, we found that this problem is well suppressed in our analysis as demonstrated below.

\begin{figure}
    \centering
    \includegraphics[width=0.9\linewidth]{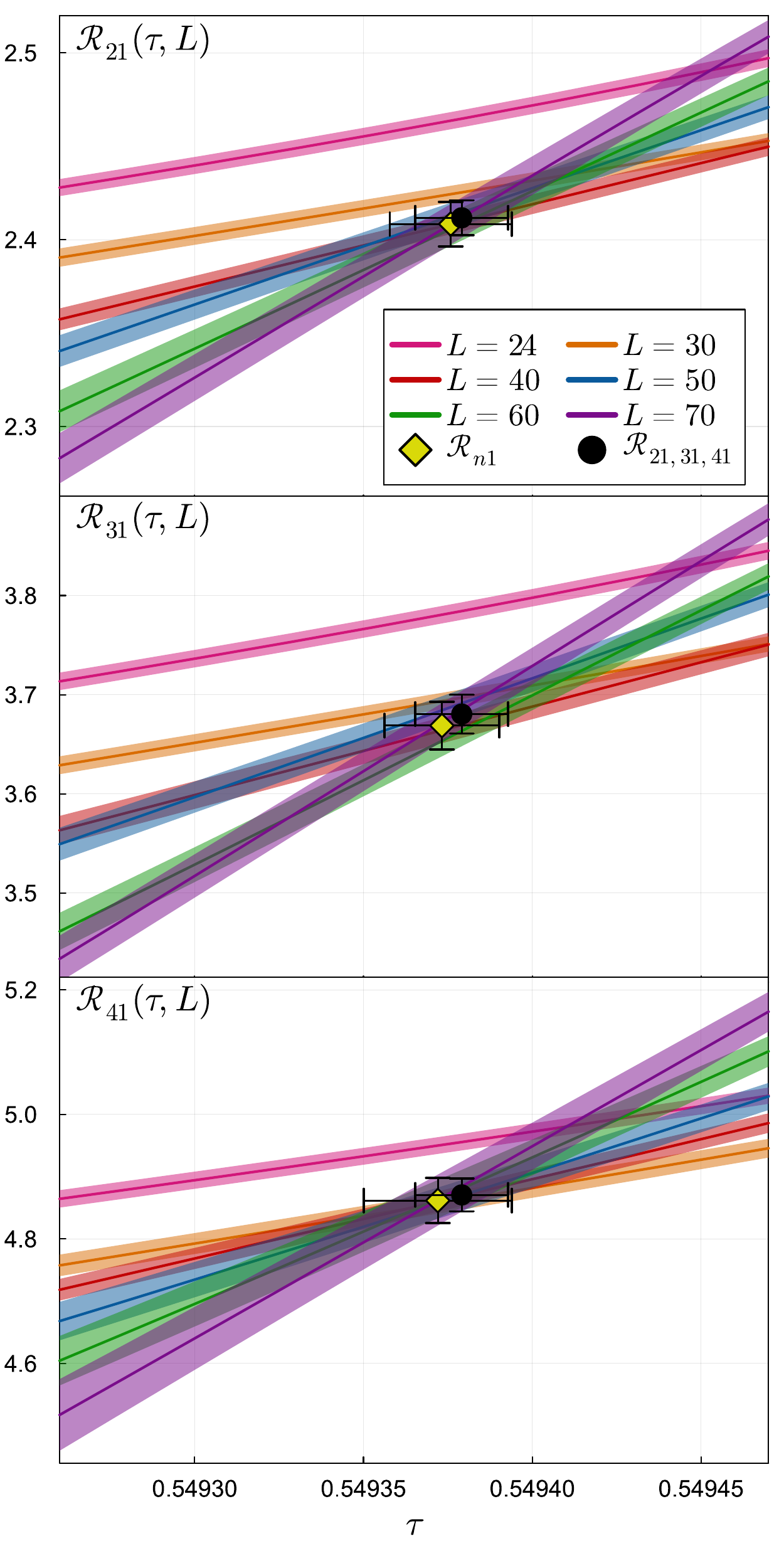}
    \caption{${\cal R}_{n1}(\tau,L)$ for $n=2,3,4$ and various $L$. The diamond and circle markers are the fit results for single and all ratios, respectively.}
    \label{fig:R}
\end{figure}

%In Fig.~\ref{fig:LYZ}, are the LYZ for $n=1,2,3$ 

\begin{table}[b]
    \caption{Fit results of the CP parameters and $\chi^2/{\rm d.o.f}$. }
    \label{tab:result}
    \centering
    \begin{tabular}{lcccc}
    \hline\hline
    \rule{0pt}{2.5ex}fit data & $\tau_{\rm c}$ & $y_t$ & $r_{n1}$ or $b_4$ & $\chi^2/\rm{d.o.f}$
    \\
    \hline
    ${\cal R}_{21}$ & 
   0.549375(18) & 1.53(19) & 2.408(12) & 0.38 \\
    ${\cal R}_{31}$ & 
   0.549373(17) & 1.66(19) & 3.669(24) & 0.38 \\
    ${\cal R}_{41}$ & 
   0.549372(22) & 1.71(21) & 4.861(36) & 0.55 \\
    ${\cal R}_{32}$& 
   0.549381(48) & 2.04(58)& 1.5257(62) & 0.40 \\
    ${\cal R}_{42}$ & 
   0.549395(43) & 2.32(60)& 2.0249(90) & 0.62 \\
    ${\cal R}_{43}$ & 
   0.549418(103) & 2.36(150)& 1.3258(55) & 0.91 \\
    ${\cal R}_{21,31,41}$ & 
   0.549379(14) & 1.70(16) &   $\cdots$      & 0.56 \\
    ${\cal B}_4$ &
   0.549382(11) & 1.63(13) & 1.614(8) & 0.69\\\hline\hline
    \end{tabular}
\end{table}

In Fig.~\ref{fig:R}, we show the ratios ${\cal R}_{21}(\tau,L)$, ${\cal R}_{31}(\tau,L)$, and ${\cal R}_{41}(\tau,L)$ as functions of $\tau$ for various $L$, where the shaded bands represent statistical errors estimated by the j%J
ackknife method with 20 bins on all configurations. The figure shows that the results for various $L$ intersect at almost a common point in all ratios as anticipated from Eq.~\eqref{eq:Rlinear2}, except for the results for $L=24,30$ having clear deviations that would be attributed to the finite-volume effect.

To obtain the critical value $\tau=\tau_{\rm c}$ at the CP, we performed the four-parameter chi-square fits to the data of ${\cal R}_{nm}(\tau,L)$ for $L\ge40$ ($12$ data points in total) with an ansatz ${\cal R}_{nm}(\tau,L)=r+c(\tau-\tau_{\rm c})L^{y_t}$ with $r$, $c$, $\tau_{\rm c}$, and $y_t$ being the fitting parameters. Effects of the second bracket in Eq.~\eqref{eq:Rlinear2} are neglected since no deviation of the intersection point is visible for $L\ge40$ in Fig.~\ref{fig:R}. The fit results are shown by the diamonds in Fig.~\ref{fig:R} for $m=1$ and summarized in Table~\ref{tab:result}. The table shows that %the results for $n=2,3,4$ are consistent with each other. 
these results are consistent with each other, while smaller $m$ tends to give better statistics with fixed $n$.

To fully make use of the information for $n=2,3,4$ to determine $\tau_{\rm c}$, we also performed the eight-parameter correlated fit to ${\cal R}_{21}(\tau,L)$, ${\cal R}_{31}(\tau,L)$, ${\cal R}_{41}(\tau,L)$ with the common $\tau_{\rm c}$ and $y_t$. The results are shown in Fig.~\ref{fig:R} by the circles and in Table~\ref{tab:result} (the row labeled ${\cal R}_{21,31,41}$). One finds that this analysis gives a better statistics than the above ones. All the fits give reasonable $\chi^2/{\rm dof}$ as in the far-right columns in Table~\ref{tab:result}.

\begin{figure}
    \centering    \includegraphics[width=0.9\linewidth]{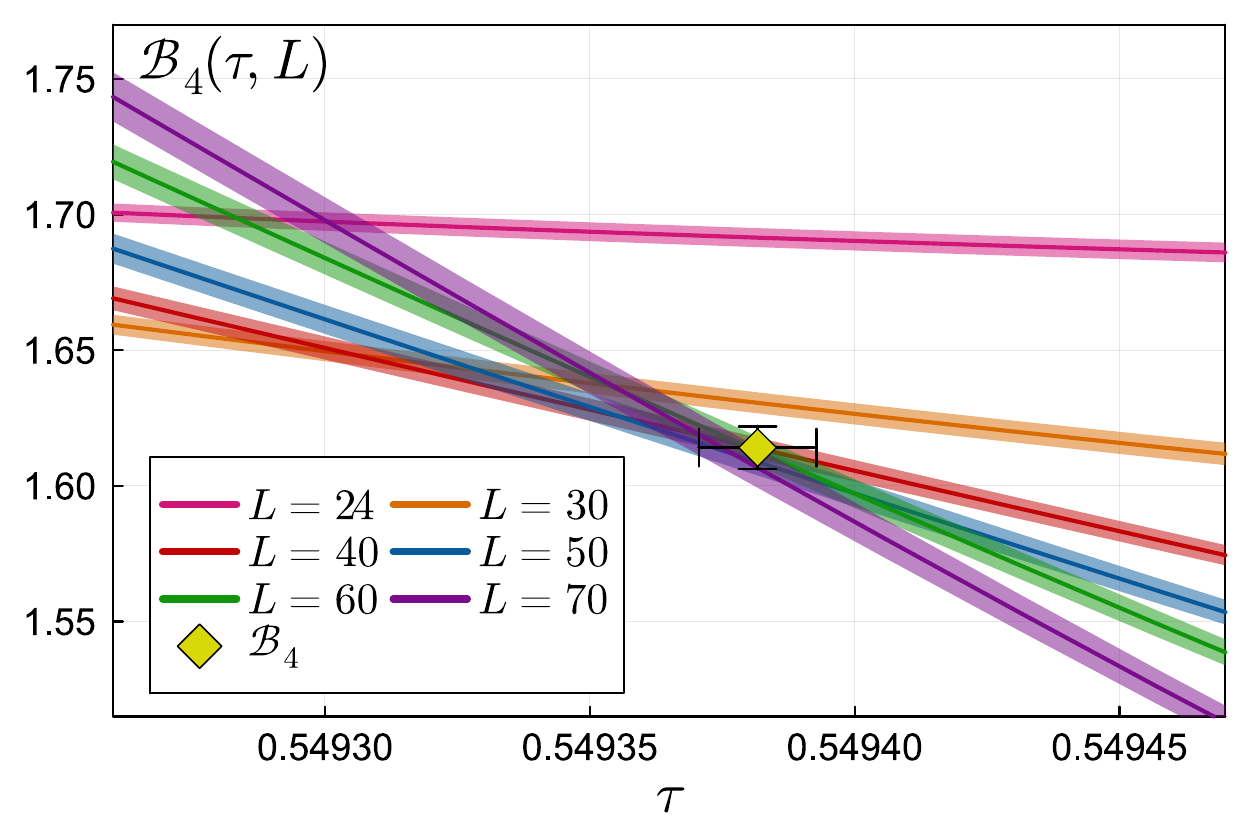}
    \caption{${\cal B}_4(\tau,L)$ for various $L$. The meanings of symbols are the same as Fig.~\ref{fig:R}.}
    \label{fig:B4}
\end{figure}

To compare these results with the Binder-cumulant analysis, in Fig.~\ref{fig:B4} we show the behavior of ${\cal B}_4(\tau,L)$ obtained on the same configurations. The fit result with the same procedure as above is shown by the diamond in the figure and in Table~\ref{tab:result}, which is consistent with the result in Ref.~\cite{Karsch:2000xv}. The resulting values of $\tau_{\rm c}$ and $y_t$ are consistent with those obtained from the LYZ ratio, while the statistical errors are almost the same in both methods.

\Sec{Summary and outlook}
In this Letter, we discussed the FSS of LYZ and showed that the intersection point of their ratios, Eq.~\eqref{eq:Rnm} or Eq.~\eqref{eq:Rlinear3}, for various spatial volumes indicates a CP in general systems. This property can be used for numerical searches of CPs as an independent method from the conventional ones based on susceptibilities. Compared to the Binder-cumulant method~\cite{Binder:1981sa}, this method is advantageous in suppressing the finite-volume effects arising from $a_{12}\ne0$. We applied the method to the numerical analysis of the CP in the three-state Potts model. Whereas we assumed the CP in the 3D-Ising universality class throughout the Letter for a simple presentation, this method, of course, can be %adapted 
extended to CPs in other universality classes.

In our numerical study, we limited our analysis to the LYZ with $n\le4$. However, one can use the LYZ for yet larger $n$, which will act to improve the statistics. The LYZ can also be used to determine $\xi_{\rm c}$ and the matrix $A$, where better control of the finite-volume effects and statistics will be realized by combined uses of Eqs.~\eqref{eq:Rexi} and~\eqref{eq:Imxi} for various $n$. The precise measurements of $r_{nm}$ in individual universality classes and the violation of the FSS~\eqref{eq:FSS} in specific models~\cite{Ferrenberg:2018zst,Kaupuzs:2022wtd}, as well as the application of the method to CPs in various systems, such as the QCD critical point in lattice simulations, are other important future studies.

\Sec{Acknowledgement}
We thank Shinji Ejiri for discussions in the early stage of this study. The authors also thank Koji Hukushima and Christian Schmidt for useful discussions.
This work was supported in part by JSPS KAKENHI (Nos.~JP19H05598, JP22K03593, JP22K03619, JP23H04507, JP24K07049), the Center for Gravitational Physics and Quantum Information (CGPQI) at Yukawa Institute for Theoretical Physics, the Research proposal-based use at the Cybermedia Center, Osaka University, and the Multidisciplinary Cooperative Research Program of the Center for Computational Sciences, University of Tsukuba.

\bibliography{bibliography}
\clearpage
\end{document}